\documentstyle[pre,preprint,aps]{revtex}
\tightenlines

\begin{document}
\draft

\title{
Stochastic Resonance of Ensemble Neurons for \\
Transient Spike Trains: A Wavelet Analysis
}
\author{
Hideo Hasegawa
\footnote{E-mail:  hasegawa@u-gakugei.ac.jp}
}
\address{
Department of Physics, Tokyo Gakugei University,
Koganei, Tokyo 184-8501, Japan
}
\date{\today}
\maketitle
\begin{abstract}
By using the wavelet transformation (WT),
we have analyzed the response
of an ensemble of $N$ (=1, 10, 100 and 500) 
Hodgkin-Huxley (HH) neurons 
to {\it transient} $M$-pulse spike trains ($M=1-3$)
with independent Gaussian noises.
The cross-correlation between the input and output 
signals is expressed
in terms of the WT expansion coefficients.
The signal-to-noise ratio (SNR)
is evaluated by using the {\it denoising} method within the WT,
by which the noise contribution is extracted 
from output signals.
Although the response of a single ($N=1$) neuron to 
sub-threshold transient signals with noises
is quite unreliable, the transmission fidelity
assessed by the cross-correlation and SNR
is shown to be much improved
by increasing the value of $N$:
a population of neurons play an indispensable role in 
the stochastic resonance (SR) for transient spike inputs.
It is also shown that in a large-scale ensemble, the transmission fidelity
for supra-threshold transient spikes is not significantly degraded
by a weak noise which is responsible to SR for sub-threshold inputs. 
\end{abstract}

\noindent
\vspace{0.5cm}
%
\section{INTRODUCTION}


In recent years much
studies have been made for the stochastic resonance (SR),
in which weak input signals are enhanced by background
noises \cite{Gammai98}-\cite{Anish99}.
This paradoxical SR phenomenon was first discovered
in the context of climate dynamics, and it is now
reported in many non-linear systems such as
electric circuits, ring lasers, semiconductor devices
and neurons. 

For single neurons, SR has been studied by 
using various theoretical models such as
the integrate-and-fire (IF) 
model \cite{Bulsara96}-\cite{Shimokawa99a},
the FitzHough-Nagumo (FN) model \cite{Longtin93}-\cite{Longtin94}
and the Hodgkin-Huxley (HH) model \cite{Lee99}.
In these studies, a weak periodic (sinusoidal) signal
is applied to the neuron, and 
it has been reported that the peak height of the 
interspike-interval (ISI) distribution \cite{Bulsara96}-\cite{Longtin93}
or the signal-to-noise ratio (SNR) of 
output signals \cite{Wiesenfeld94}-\cite{Lee99} 
shows the maximum when the noise intensity is changed.

SR in coupled or ensemble neurons has been also investigated
by using the IF model \cite{Shimokawa99b}-\cite{Lindner01}, 
FN model \cite{Collins95a}-\cite{Stocks01} 
and HH model \cite{Pei96a}-\cite{Liu01}.
The transmission
fidelity is examined by calculating various quantities:
the peak-to-width ratio
of output signals 
\cite{Collins95a}\cite{Shimokawa99b}\cite{Pei96a}\cite{Tanabe99}, 
the cross-correlation between input 
and output signals \cite{Shimokawa99b}\cite{Liu01}, 
SNR \cite{Shimokawa99b}\cite{Kanamaru01}\cite{Liu01}, 
and the mutual information \cite{Stocks01}.
One or some of these quantities 
has been shown to take a maximum
as functions of the noise intensity and
the coupling strength.
Collins, Chow and Imhoff \cite{Collins95a} have pointed out
that SR of ensemble neurons is improved
as the size of an ensemble is increased.
Some physiological experiments support
SR in real, biological 
systems of crayfish \cite{Douglass93}\cite{Pei96b},
cricket \cite{Levins96}
and rat \cite{Gluckman96}\cite{Nozaki99}.

SR studies mentioned above are motivated from the fact that
peripheral sensory neurons play a role of transducers,
receiving analog stimula and emitting spikes.
In central neural systems, however, cortical neurons 
play a role of data-processors,
receiving and transmitting 
spike trains.
The possibility of SR in the spike transmission has been 
reported \cite{Chapeau96}-\cite{Mato98}.
The response to periodic coherent spike-train inputs
has been shown to be enhanced by an addition of
weak spike-train noises whose interspike intervals (ISIs) 
have the Poison or gamma distribution.

It should be stressed that these theoretical studies on 
SR in neural systems have been performed mostly
for stationary analog (or spike-train) signals
although they are periodic or aperiodic \cite{Collins95b}.
There has been few theoretical studies on SR
for non-stationary signals.
Fakir \cite{Fakir98} has discussed SR
for non-stationary analog inputs with finite duration
by calculating the cross-correlation.
By applying a single impulse, 
Pei, Wilkinson and Moss \cite{Pei96a}
have demonstrated that the spike-timing precision
is improved by noises in an ensemble of 1000 HH neurons.
One of the reason why 
SR study for stationary signals is dominant,
is mainly due to the fact that 
the stationary signals can be easily analyzed by
the Fourier transformation (FT) with which, for example,
the SNR is evaluated from FT spectra of output signals.
The FT requires that a signal to be examined is stationary, 
not giving the time evolution of the frequency pattern. 
Actual biological signals are,
however, not necessarily stationary.
It has been reported that
spike signals in cortical neurons are generally not stationary, 
rather they are transient signals or bursts \cite{Traub99}, 
whereas periodic spikes are found in systems 
such as auditory systems of owl \cite{Sullivan98}
and the electrosensory system of electric fish \cite{Rose85}.

The limitation of the FT analysis can be partly resolved by using 
the short-time Fourier transformation (STFT).  Assuming that the signal
is quasi-stationary in the narrow time period,
the FT is applied with time-evolving narrow windows. 
Then STFT yields the time evolution of the frequency spectrum.
The STFT, however, has a critical limitation violating the
{\it uncertainty principle}, which asserts that if the window is too
narrow, the frequency resolution will be poor whereas
if the window is too wide, the time resolution will be less precise.
This limitation becomes serious for signals with much transient 
components, like spike signals.

The disadvantage of the STFT is overcome 
in the wavelet transformation (WT) \cite{Astaf96}.
In contrast to the FT, the WT offers the two-dimensional expansion for
a time-dependent signal with the scale and translation parameters
which are interpreted physically as the inverse of frequency and
time, respectively. 
As a basis of the WT, we employ the {\it mother wavelet}
which is localized in both frequency and time domains.
The WT expansion is carried out in terms of a family of wavelets 
which is made by dilation and translation of the mother wavelet.
The time evolution of frequency pattern can be followed with an optimal
time-frequency resolution.

The WT appears to be an ideal tool for analyzing signals of 
a non-stationary nature.
In recent years the WT has been applied to an analysis of 
biological signals \cite{Samar99},
such as electoencephalographic (EEG) 
waves \cite{Blanco96}-\cite{Rosso01},
and spikes \cite{Hulata00}-\cite{Hasegawa01}.
EEG is a reflection of the activity of ensembles of neurons 
producing oscillations. By using the WT, we
can obtain the time-dependent decomposition of
EEG signals to $\delta$ (0.3-3.5 Hz), 
$\theta$ (3.5-7.5 Hz), $\alpha$ (7.5-12.5 Hz),
$\beta$ (12.5-30.0 Hz) and $\gamma$ (30-70 Hz) 
components \cite{Blanco96}-\cite{Rosso01}.
It has been shown that the WT is a powerful tool to the spike sorting
in which coherent signals of a single target neuron 
are extracted from mixture
of response signals \cite{Hulata00}-\cite{Zour97}.
Quite recently Hasegawa \cite{Hasegawa01} has 
made an analysis of transient spike-train
signals of a HH neuron with the use of WT, calculating the 
energy distribution and Shanon entropy.

It is interesting to analyze the response of ensemble neurons
to {\it transient} spike inputs in noisy environment
by using the WT.
There are several sources of noises:
(i) cells in sensory neurons are exposed to
noises arising from the outer world,
(ii) ion channels of the membrane of neurons are known
to be stochastic \cite{Destexhe98},
(iii) the synaptic transmission yields noises originating 
from random fluctuations of the synaptic vesicle 
release rate \cite{Smith98}, and 
(iv) synaptic inputs include leaked currents from
neighboring neurons \cite{Shadlen94}.
Most of existing studies on SR adopt the Gaussian noises,
taking account of the items (i)-(iii).
Simulating the noise of the item (iv), 
Refs.\cite{Chapeau96}-\cite{Mato98}
include spike-train noises whose ISIs have the Poisson
or gamma distribution.
In this study we take into account Gaussian noises;
SR for spike-train inputs 
with added spike-train noises will be discussed
in a separate paper \cite{Hasegawa02}.

We assume an ensemble of Hodgkin-Huxley (HH) neurons to
receive transient spike trains with 
independent Gaussian noises.
The HH neurons model is adopted because 
it is known to be the most realistic among
theoretical models \cite{Hodgkin52}.
The signal transmission is assessed by the cross-correlation between
input and output signals and SNR, which are expressed
in terms of WT expansion coefficients.
In calculating the SNR, we adopt the denoising technique within
the WT method \cite{Bartnik92}-\cite{Quiroga00}, 
by which the noise contribution is extracted from output signals.

Our paper is organized as follows:
In Sec. IIA, an adopted model
for an ensemble of $N$-unit HH neurons is described, and in Sec. IIB 
the WT is briefly discussed.
We present the calculated results for $M=1$ pulse
train in Sec. IIIA, and the results for
spike trains with $M=2$ and 3 are discussed in Sec.IIB.
The final Sec. IV is devoted to conclusion and discussion.

\section{CALCULATION METHODS}
\subsection{Ensemble Neuron Model}


We assume 
a network consisting of
$N$-unit HH neurons which receive
the same spike trains 
but independent Gaussian white noises through
excitatory synapses.
Spikes emitted by the ensemble neurons are collected
by a summing neuron.
A similar model was previously adopted by several
authors studying SR for {analog}
signals \cite{Collins95a}\cite{Pei96a}\cite{Tanabe99}.
An input signal in this paper 
is a transient spike train 
consisting of $M$ impulses ($M=1-3$).
Dynamics of the membrane potential $V_{i}$ of
the HH neuron {\it i}
is described by the non-linear differential
equations given by 

\begin{equation}
\bar{C} \:d V_{i}(t)/d t = -I_{i}^{\rm ion}  
+ I_{i}^{\rm ps} + I_{i}^{\rm n},
\;\;\;\;\;\;\;\;\;\;\;\;\;\; 
\mbox{(for $1 \leq i \leq N$)} \\
\end{equation}
where 
$\bar{C} = 1 \; \mu {\rm F/cm}^2$ is the capacity of the membrane.
The first term $I_{i}^{\rm ion}$ of Eq.(1) denotes the ion current
given by
\begin{equation}
I_{i}^{\rm ion} 
= g_{\rm Na} m_{i}^3 h_{i} (V_{i} - V_{\rm Na})
+ g_{\rm K} n_{i}^4 (V_{i} - V_{\rm K}) 
+ g_{\rm L} (V_{i} - V_{\rm L}),
\end{equation}
where
the maximum values of conductivities 
of Na and K channels and leakage are
$g_{\rm Na} = 120 \; {\rm mS/cm}^2$, 
$g_{\rm K} = 36 \; {\rm mS/cm}^2$ and
$g_{\rm L} = 0.3 \; {\rm mS/cm}^2$, respectively; 
the respective reversal potentials are   
$V_{\rm Na} = 50$ mV, $V_{\rm K} = -77$ mV and 
$V_{\rm L} = -54.5 $ mV.
Dynamics of the gating variables of Na and
K channels, $m_{i}, h_{i}$ and $n_{i}$,
are described by the ordinary differential equations,
whose details have been given elsewhere 
\cite{Hodgkin52}\cite{Hasegawa00}.

The second term $I_{i}^{\rm ps}$ in Eq.(1) denotes 
the postsynaptic current given by
\begin{equation}
I_{i}^{\rm ps} = \; \sum_{m=1}^M 
g_{s}\:(V_a - V_s) \:\alpha(t-t_{{\rm i}m}), 
\end{equation}
which is induced by an input spike with the magnitude $V_{a}$ given by
\begin{equation}
U_{i}(t) = V_{a} \;\sum_{m=1}^{M}\; \delta(t-t_{{\rm i}m}),
\end{equation}
with the alpha function $\alpha(t)$:
\begin{equation}
\alpha(t) = (t/\tau_{\rm s}) \; e^{-t/\tau_{\rm s}} \:  \Theta(t).
\end{equation}
In Eqs.(3)-(5) $t_{{\rm i}m}$ is the $m$-th firing time of the input,
the Heaviside function is defined by
$\Theta (t)=1$ for $x \geq 0$ and 0 for $x < 0$,
and $g_{s}$, $V_{\rm s}$ and $\tau_{\rm s}$ stand 
for the conductance, reversal potential and time constant,
respectively, of the synapse.

The third term $I_{i}^{\rm n}$ in Eq.(1) denotes the Gaussian noise
given by
\begin{equation}
< \overline{I_{i}^{n}(t)}> = 0, 
\end{equation}
\begin{equation}
<\overline{I_{i}^{n}(t) \:I_{\ell}^{n}(t')}> 
= 2 D \:\delta_{i\ell} \:\delta(t-t'),
\end{equation}
where the overline $\overline{ X }$ and 
the bracket $ < X >$ denote the temporal and
spatial averages, respectively, and 
$D$ the intensity of white noises. 

The output spike of the neuron $i$ in an ensemble is given by
\begin{equation}
U_{{\rm o}i}(t) = V_{a} \;\sum_{n}\; \delta(t-t_{{\rm o}in}),
\end{equation}
in a similar form as an input spike [Eq.(4)],
where $t_{{\rm o}in}$ is the $n$-th firing time when
$V_{i}(t)$ crosses $V_{z}=0$ mV from below.

We should remark that our model given by Eqs.(1)-(7) does 
not include couplings among ensemble neurons.
This is in contrast with some works on ensemble neurons
\cite{Kanamaru01}\cite{Stocks01}\cite{Wang00}\cite{Liu01}
where introduced couplings among 
neurons play an important role in SR besides noises,
related discussion being given in Sec. IV.

Differential equations given by Eqs.(1)-(7)
are solved by the forth-order Runge-Kutta method
by the integration time step of 0.01 ms
with double precision.
Some results are examined by using
the exponential method.
The initial conditions for the variables are given by
$V_{i}(t)$= -65 mV, $m_{i}(t)$=0.0526,  
$h_{i}(t)$=0.600, 
$n_{i}(t)$=0.313
at $t=0$,
which are the rest-state solution of a single
HH neuron.
Hereafter time, voltage, conductance,
current, and $D$
are expressed in units of ms, mV, ${\rm mS/cm}^2$, $\mu {\rm A/cm}^2$,
and $\mu {\rm A}^2/cm^4$, respectively.
We have adopted parameters of
$V_a=30$, $V_c=-50$,
and $\tau_{\rm s} = \tau_{n}=2$. 
Adopted values of $g_{s}$,
$D$, $M$ and $N$ will be described shortly.

\subsection{Wavelet Analysis}

There are two types of WTs: one is the continuous wavelet transformation (CWT)
and the other the discrete wavelet transformation (DWT). In the former the parameters
denoting the scale and trasnlation are continuous variables while in the latter
they are discrete variables.

The CWT for a given
regular function
$f(t)$ is defined by
\begin{eqnarray}
c(a, b) = \int {\rm d}t  \; \psi_{a b}^{*}(t) \;f(t) 
\equiv <\psi_{a b}(t),\: f(t)>, 
\end{eqnarray}
with a family of wavelets $\psi_{a b}(t)$ generated by
\begin{eqnarray}
\psi_{a b}(t) = \mid a \mid^{-1/2} \psi(\frac{t-b}{a}),
\end{eqnarray}
where $\psi(t)$ is the {\it mother wavelet}, 
the star denotes the complex conjugate, and
$a$ and $b$ express the scale change and translation, 
respectively, and they
physically stand for the inverse of
the frequency and the time.
Then the CWT transforms the time-dependent function $f(t)$
into the frequency- and time-dependent function $c(a,b)$. 
The mother wavelet is a smooth function with good localization
in both frequency and time spaces.
A wavelet family given by Eq.(10) plays a role of elementary
function, representing the function $f(t)$ as a superposition
of wavelets $\psi_{a b}(t)$.

The {\it inverse} of the wavelet transformation may be given by
\begin{equation}
f(t) = C_{\psi}^{-1}  \int \: \frac{{\rm d}a}{a^2} \int {\rm d}b \;
c(a, b) \;\psi_{a b}(t),
\end{equation}
when the mother wavelet satisfies the following two conditions:

\noindent
(i) the admissibility 
condition given by
\begin{equation}
0 < C_{\psi} < \infty,
\end{equation}
with
\begin{equation}
C_{\psi} = 
\int_{-\infty}^{\infty} {\rm d}\omega \mid \hat{\Psi}(\omega) \mid^2/
\mid \omega \mid, 
\end{equation}
where $\hat{\Psi}(\omega) $ is the Fourier
transformation of $\psi(t)$,
and 

\noindent
(ii) the zero mean of the mother wavelet:
\begin{equation}
\int_{-\infty}^{\infty}  {\rm d}t \; \psi(t) = \hat{\Psi}(0) = 0.
\end{equation}

\vspace{0.5cm}


On the contrary, the DWT is defined
for {\it discrete} values of $a=2^{j}$ and $b=2^{j}k$ 
($j,k$: integers) as
\begin{eqnarray}
c_{jk} \equiv c(2^{j}, 2^{j}k) = <\psi_{jk}(t),\; f(t)>, 
\end{eqnarray}
with
\begin{eqnarray}
\psi_{jk}(t) = 2^{-j/2} \psi(2^{-j} t - k).
\end{eqnarray}
The ortho-normal condition for the wavelet functions is
given by
\begin{eqnarray}
<\psi_{jk}(t), \: \psi_{j'k'}(t)> 
= \delta_{jj'}\:\delta_{kk'},
\end{eqnarray}
which leads to the inverse DWT:
\begin{eqnarray}
f(t) = \sum_{j} \; \sum_{k} \;c_{jk} \;\psi_{jk}(t).
\end{eqnarray}

In the multiresolution analysis (MRA) 
of the DWT, we introduce a scaling function
$\phi(t)$, which satisfies the recurrent relation with $2K$
masking coefficients, $h_{k}$, given by
\begin{eqnarray}
\phi(t) = \surd 2\sum_{k=0}^{2K-1} h_k \;\phi(2t- k),
\end{eqnarray}
with the normalization condition for $\phi(t)$ given by
\begin{eqnarray}
\int {\rm dt} \; \phi(t) = 1.
\end{eqnarray}
A family of wavelet functions is generated by
\begin{eqnarray}
\psi(t) = \surd 2\sum_{k=0}^{2K-1} (-1)^k h_{2K-1-k} \;\phi(2t- k).
\end{eqnarray}
The scaling and wavelet functions satisfy the ortho-normal
relations:
\begin{eqnarray}
<\phi(t), \phi(t-m)> &=& \delta_{m0}, \\
<\psi(t), \psi(t-m)> &=& \delta_{m0}, \\
<\phi(t), \psi(t-m)> &=& 0.
\end{eqnarray}
A set of masking coefficients $h_j$ 
is chosen so as to satisfy the conditions shown above.

The simplest wavelet function for $K=1$ is the Harr wavelet
for which we get $h_0 = h_1 = 1/\surd 2$, and 
\begin{eqnarray}
\psi_{\rm H}(t) &=& 1, \;\;\mbox{for $0 \leq t < 1/2$} \nonumber \\
&=& -1, \;\;\mbox{for $1/2 \leq t < 1$} \nonumber \\
&=& 0, \;\;\mbox{otherwise}
\end{eqnarray}
In the more sophisticated wavelets like
the Daubechies wavelet, an additional condition given by
\begin{eqnarray}
\int {\rm d}t\; t^{\ell} \;\psi(t) = 0, \;\;\; 
\mbox{for $\ell=0,1,2,3.....L-1$}
\end{eqnarray}
is imposed for the smoothness of the wavelet function.
Furthermore,
in the Coiflet wavelet, for example, a similar smoothing 
condition is imposed also for the scaling function as
\begin{eqnarray}
\int {\rm d}t\; t^{\ell} \;\phi(t) = 0, \;\;\; 
\mbox{for $\ell=1,2,3.....L'-1$}
\end{eqnarray}

Once WT coefficients are obtained, we can calculate
various quantities such as auto- and cross-correlations
and SNR, as will be discussed shortly.
In principle the expansion coefficients $c_{jk}$ in DWT 
may be calculated 
by using Eqs.(15) and (16) for a given function $f(t)$ and
an adopted mother wavelet $\psi(t)$. 
This integration is, however, inconvenient,
and in an actual fast wavelet transformation, 
the expansion coefficients are obtained 
by a matrix multiplication with the use of the iterative formulae
given by the masking coefficients and 
expansion coefficients of the neighboring levels of indices,
$j$ and $k$ \cite{Astaf96}. 

One of the advantages of the WT over FT 
is that we can choose a proper mother wavelet
among many mother wavelets, depending on 
a signal to be analyzed.
Among many candidates of mother wavelets, we have
adopted the Coiflet with level 3, compromising
the accuracy and the computational effort.
The WT has been performed by using
the MATLAB wavelet tool box.

\section{CALCULATED RESULTS}


\subsection{Input Pulses with $M=1$}

Firstly we discuss the case in which
ensemble HH neurons receive a common single ($M=1$) impulse.
When input synaptic strength is small:
$g_s < g_{th}$, no neurons fire in the noise-free case,
while it is sufficiently large ($g_s \geq g_{th}$) neurons fire,
where $g_{th}=0.091$ is the threshold value.
For a while, we will discuss the sub-threshold case 
of $g_{s}=0.06 < g_{th}$ with $N=500$.
The $M=1$ input pulse is applied at $t=100$ ms,
as shown in Fig. 1(a). Figure 1(b) shows
the time dependence of the postsynaptic current
$I_{i}=I_{i}^{\rm ps}+I_{i}^{\rm n}$ of the neuron $i=1$ with added noises
of $D=0.5$.
Because neurons receive noises for $0 \leq t < 100$ ms,
the states of neurons when they receive 
input pulse are randomized \cite{Tanabe99}.
Figure 1(c) shows the time dependence of the membrane
potential $V_1$ of the neuron 1, which fires with 
a delay of about 6 ms.  This delay is much larger
than the conventional value of 2-3 ms for the supra-threshold inputs
because the integration of marginal inputs at synapses
needs a fairly a long period before firing \cite{Hasegawa00}.
Firings in ensemble 500 neurons for $D=0.5$ are depicted by 
raster in Fig. 2(a).
We note that neurons fire not only by input pulses plus noises
but also spuriously by noises only.
When the noise intensity is increased to $D=1.0$,
spurious firings are much increased, as Fig. 2(b) shows.

We assume that information is carried
by firing times of spikes but not by
details of their shapes.
In order to study how information is transmitted
through ensemble HH neurons with the use of the DWT,
we divide the time scale by the width of time bin
of $T_{\rm b}$ as $t = t_{\ell}= \ell \;T_{\rm b}$ ($\ell$: integer), and 
define the input and output
signals within the each time bin by
\begin{equation}
W_{\rm i}(t) = \sum_{m=1}^M \;\Theta(\mid t - t_{{\rm i}m}\mid - T_{\rm b}/2),
\end{equation}

\begin{equation}
W_{\rm o}(t) = (1/N) \:\sum_{i=1}^{N} \sum_n \;
\Theta(\mid t - t_{{\rm o}in}\mid - T_{\rm b}/2).
\end{equation}
In Eqs. (28) and (29) $\Theta(t)$ stands for 
the Heaviside function,
$W_{\rm i}(t)$ the external input signal
(without noises),
$W_{\rm o}(t)$ the output signal averaged over the ensemble neurons,
$t_{{\rm i}m}$ the $m$-th firing time of inputs, 
and $t_{{\rm o}in}$ the $n$-th
firing time of outputs of the neuron $i$ [Eq.(8)].
The time bin is chosen as $T_{\rm b}$ =5 ms in our simulations.
A single simulation has been performed for 320 (=$2^6\;T_{\rm b}$) ms.
Figures 1(d) and 1(e) show the time dependence of $W_{\rm i}(t)$
and $W_{\rm o}(t)$, respectively, for the case of $D=0.5$.
The magnitude of $W_{\rm o}(t)$ is much smaller than that of $W_{\rm i}(t)$
because only a few neurons fire among 500 neurons.
The peak position of $W_{\rm o}(t)$ is slightly shifted
compared with that of $W_{\rm i}(t)$ because of a significant 
delay of neuron firings as mentioned above.

\vspace{0.5cm}
\noindent
{\it Wavelet Transformation}
\vspace{0.5cm}

Now we apply the DWT to input  
and output signals .
By substituting $f(t)=W_{\rm i}(t)$ or
$W_{\rm o}(t)$ in Eq.(15), 
we get their WT coefficients 
given by
\begin{equation}
c_{\lambda jk} 
= \int {\rm d}t  \; \psi_{j k}^{*}(t) \;W_{\lambda}(t),
\;\;\;\;\; \mbox{($\lambda=$ i, o)}
\end{equation}
where $\psi_{jk}(t)$ is a family of wavelets generated
from the mother Coiflet wavelet [see Eq.(16)].
The uppermost frame of Fig. 3(a) expresses the input
signal $W_{\rm i}(t)$.
Note that the lower and upper horizontal scales express $b$ and
$b\;T_{b}$ (in units of ms), respectively.
Figure 3(b) shows the calculated WT coefficients
of $W_{\rm i}(t)$ which are 
plotted as a function of $b(j)=(k-0.5)\; 2^j$ 
for various $j$ values after convention.
The WT decomposition of the signal:
$f(t)=\sum_{j=1}^{5} f_j(t)$, is plotted in Fig. 3(a).
The WT coefficients of $j=1$ and 2 have large values
near $b=20$ ($b\;T_{b}=100$ ms) 
where $W_{\rm i}(t)$ has a peak.
Contributions from $j=1$ and $j=2$ are predominant 
in $W_{\rm i}(t)$.
It is noted that the WT coefficient for $j=4$
has a significant value at $b \sim 56$ far away
from $b=20$.
The WT decomposition and WT coefficients
for output signal $W_{\rm o}(t)$
for $D=0.5$ are shown in Fig. 4(a) and 4(b), respectively.
The dominant contribution arises from $j=1$
in $W_{\rm o}(t)$.
Similar plots of the WT decomposition and the WT coefficients
of the output signal $W_{\rm o}(t)$ for $D=1.0$ are presented
in Figs. 5(a) and 5(b), respectively.
As the noise intensity is increased, fine structures in
the WT coefficients appear, in particular for small $j$.

\vspace{0.5cm}
\noindent
{\it Auto- and Cross-correlations}
\vspace{0.5cm}

The auto-correlation functions for input
and output signals are defined by
\begin{eqnarray}
\Gamma_{\lambda \lambda} &=& M^{-1}\;\int {\rm d}t 
\;W_{\lambda}(t)^{*}\; W_{\lambda}(t), \\
&=& M^{-1}\;\sum_{j} \; \sum_{k} \;c_{\lambda jk}^{*}\; c_{\lambda jk},
\;\;\;\;\;\;\;\;\;\;\;\; \mbox{($\lambda$= i, o)}
\end{eqnarray}
where the ortho-normal relations of the wavelets 
given by Eqs.(22)-(24) are employed.
Similarly the cross-correlation between input and output signals 
is defined by
\begin{eqnarray}
\Gamma_{i o}(\beta) &=& M^{-1}\;\int {\rm d}t 
\;W_{i}(t)^{*}\; W_{o}(t+\beta \;T_{\rm b}),  \\ 
&=& M^{-1}\;\sum_{j} \; \sum_{k} 
\;c_{{\rm i}jk}^{*}\; c_{{\rm o}jk}(\beta),
\end{eqnarray}
where $c_{{\rm i}jk}$ and $c_{{\rm o}jk}(\beta)$ are
the expansion coefficients
of $W_{i}(t)$ and $W_{o}(t+\beta T_{\rm b})$, respectively.
The maxima in the cross-correlation and
the normalized one are given by

\begin{equation}
\Gamma = max_{\beta} [ \Gamma_{\rm io}(\beta) ], 
\end{equation}

\begin{equation}
\gamma = max_{\beta} [ \frac{\Gamma_{\rm io}(\beta)}
{\surd \Gamma_{\rm ii} \surd \Gamma_{\rm oo}} ].
\end{equation}
It is noted that for the suprashreshold inputs in the
noise-free case, we get 
$\Gamma_{\rm ii}=\Gamma_{\rm oo}=\Gamma_{\rm io}=1$ 
and then $\Gamma=\gamma=1$.

Figure 6(a) shows the $D$ dependence of $\Gamma$,
$\Gamma_{\rm oo}$ and $\gamma$ for $g=0.06$ and $N=500$.
They are zero at $D=0$ because of the
adopted sub-threshold parameters.
As increasing $D$ from zero, 
$\Gamma$ and $\Gamma_{\rm oo}$ are gradually increased.
Because of the factor of $1/\surd \Gamma_{\rm oo}$
in Eq.(36), the magnitude of $\gamma$ is
larger than those of $\Gamma$ and $\Gamma_{\rm oo}$.
We note that $\gamma$ is enhanced by weak noises and
it is decreased at larger noises, which is 
a typical SR behavior.

\vspace{0.5cm}

\noindent
{\it Signal-to-Noise Ratio}
\vspace{0.5cm}

We will evaluate the SNR by employing 
the denoising method \cite{Bartnik92}-\cite{Quiroga00}.
The key point in the denoising is how to choose which 
wavelet coefficients are correlated with the signal 
and which ones with noises.
The simple
denoising method is to neglect some DWT expansion coefficients 
when reproducing the signal by
the inverse wavelet transformation.
We get the denoising signal by the inverse WT [Eq.(18)]:
\begin{equation}
W_{\lambda}^{dn}(t) = \sum_j \;\sum_k 
c_{\lambda jk}^{dn} \;\psi_{jk}(t),
\end{equation}
with the desnoising WT coefficients $c_{jk}^{dn}$
to be chosen properly
as will be discussed below.
The simplest denoising method, for example, is to assume
that WT components for $a < a_c$ in the $(a, b)$ plane
arise from noises to set the denoising WT coefficients as
\begin{eqnarray}
c_{jk}^{dn} & = & c_{jk}, \;\;\;
\mbox{for $j \geq j_c$ ($a \geq a_c$)} \nonumber \\
& = & 0,  \;\;\;\;\; \mbox{otherwise}
\end{eqnarray}
where $j_c \;(= {\rm log}_2\; a_c)$ is the critical $j$ value \cite{Bartnik92}.

In this study we adopt a more sophisticated method,
assuming that the components for 
$b < b_{\rm L}$ or $b > b_{\rm U}$ at $a < a_c$ in the (a,b) plane
are noises to set the denoising WT coefficients as
\begin{eqnarray}
c_{jk}^{dn} & = & c_{jk}, \;\;\;
\mbox{for $j \geq j_c$ or $k_{\rm L} \leq k \leq k_{\rm U}$} \nonumber \\
& = & 0,  \;\;\;\;\; \mbox{otherwise}
\end{eqnarray}
where $j_c\;(= {\rm log}_2\; a_c)$ denotes the critical $j$ value, and 
$k_{\rm L}\;(=b_{\rm L} \; 2^{-j})$
and $k_{\rm U}\;(= b_{\rm U} \; 2^{-j})$
are the lower and upper critical $k$ values,
respectively. We get the inverse, denoising signal by using 
Eq.(37) with the denoising WT coefficients determined
by Eq.(39).

From the above consideration, we may define
the signal and noise components by
\begin{eqnarray}
A_{s} &=& \int {\rm d}t 
\;W_{\rm o}^{dn}(t)^{*}\;W_{\rm o}^{dn}(t),   \\ 
&=& \sum_j \sum_k \mid c_{{\rm o} jk}^{dn} \mid^2,
\end{eqnarray}
\begin{eqnarray}
A_{n} &=& \int {\rm d}t 
\;[W_{\rm o}(t)^{*}
\;W_{\rm o}(t) - W_{\rm o}^{dn}(t)^{*}
\;W_{\rm o}^{dn}(t)],  \\ 
&=& \sum_j \sum_k 
(\mid c_{{\rm o} jk} \mid^2 - \mid c_{{\rm o} jk}^{dn} \mid^2),
\end{eqnarray}
The SNR is defined by
\begin{equation}
SNR = 10\;{\rm log}_{10} (A_{s}/A_{n})  \;\;\; \mbox{(dB)}.
\end{equation}

In the present case we can fortunately 
obtain the WT coefficients for {\it ideal} case
of noise-free and supra-threshold inputs.
We then properly determine the denoising parameters of
$j_c$, $b_L$ and $b_U$.
From the observation of the WT coefficients 
for the ideal case, which is not shown here
but is not dissimilar to those shown in
Figs. 3(a) and 4(a), we assume that
the upper and lower bounds, 
may be chosen as
\begin{equation}
b_{\rm L}=t_{{\rm o}1}/T_{\rm b}-\delta b, \;\;\;\;
b_{\rm U}=t_{{\rm o}M}/T_{\rm b}+\delta b,
\end{equation}
where $ t_{{\rm o}1}$ ($ t_{{\rm o}M}$) are 
the first ($M$-th) firing times of
output signals
in the ideal case of noise-free and supra-threshold inputs, 
and $\delta b$ denotes the marginal distance from
the $b$ values expected to be responsible 
to the signal transmission.

The denoising WT coefficients $c_{jk}^{\rm dn}$ of
$W_{\rm o}$ for $D=0.5$ and $D=1.0$ are denoted
by dashed bars in Fig. 4(b) and 5(b), respectively.
Note that the positions of dashed bars are slightly shifted 
leftward to avoid the superposition with the solid bars
representing the original WT coefficients.
In the case of $D=0.5$, the denoising WT coefficients
are almost the same as the original ones.
The inverse WT signal of $W_{\rm o}^{\rm dn}$, which is given by Eq.(37) 
and which is shown by the dashed curve in Fig. 4(a),
is almost the same as the original signal shown by the solid curve. 
On the contrary, in the case of $D=1.0$,
the denoising WT coefficients of $W_{\rm o}$ do not include
the terms far away from $b=20$ for $j < j_c$ 
in the original WT coefficients. 
The inverse WT signal of $W_{\rm o}^{dn}$, 
shown by the dashed curve in Fig. 5(a), 
has less amount of ripples than the original
signal shown by the solid curve.

Figure 7(a) and 7(b) show the calculated SNR of $W_{\rm o}$
as a function of $\delta b$ for various 
$j_c$ values with $D=0.5$ and $D=1.0$, respectively.
We note that the value of SNR is rather insensitive
to a choice of the parameters of
$\delta b$ and $j_{c}$.
Then we have decided to adopt
$\delta b=5$ and
$j_{c}=3$ for our simulations.

Figure 6(b) shows the $D$ dependence of 
SNR calculated for $g=0.06$ and $N=500$. 
We note that SNR shows a typical SR behavior:
a rapid rise to a clear peak and a slow decrease
for larger value of $D$.

\vspace{0.5cm}

So far our simulation has been made
for the fixed parameters of $g_s=0.06$ and $N=500$,
which will be changed in the followings. 
Figure 8(a) and 8(b) show the $D$ dependence
of the cross-correlation $\gamma$ and SNR, 
respectively, calculated for
$N=$ 1, 10 100 and 500.
The results for $N=1$ and 10 are averaged values
of ten and hundred trials, respectively,
their root-mean-square (RMS) values being shown by
error bars in Fig. 8(a) and 8(b).
The results of $N=500$ show the typical
SR behavior as mentioned before [Figs. 6(a) and 6(b)].
On the contrary, 
SR effect for a single ($N=1$) is marginally realized 
in SNR but not in $\gamma$. 
Large error bars for the results of $N=1$ implies
that the reliability of information transmission
is very low in the sub-threshold condition \cite{Mainen95}.
This is clearly seen in Fig. 9(a) and 9(b)
where $\gamma$ and SNR for $N=1$ are
plotted against the trial number of 100 runs.
The signal can be transmitted
only when the signal fortunately coincides
with noises and the signal plus noise crosses the threshold level.
Then only 13 among 100 trials are succeeded
in the transmission of $M=1$ inputs through a neuron.
We note in Figs. 8(a) and 8(b) that
as the size of the network is much increased,
the peak values of $\gamma$ and SNR are much enhanced
and the SR behavior becomes more evident.
Figures 8(a) and 8(b) demonstrate that the ensemble of neurons
play an {\it indispensable} role for information
transmission of transient spike signals.
This is consistent with the results of
Collins  {\it et al.} \cite{Collins95a} 
and Pei {\it et al.} \cite{Pei96a},
who have pointed out the improvement of the
information transmission by increasing the
size of the network. 

Next we change the value of $g_s$, the strength of input synapses.
The $g_s$ dependence of 
$\gamma$ and SNR for $N=500$ neurons
is shown in Figs. 10(a) and 10(b), respectively,
where values of
$D=0.0$, 0.5 and 1.0 are employed.
Note that in the noise-free case ($D=0$), we get $\gamma=1$
and SNR=$\infty$
for $g_s \geq g_{\rm th}$ but
$\gamma$ = 0 and SNR=$-\infty$ for $g_s < g_{\rm th}$,
as shown by dotted curves in Figs. 10(a) and 10(b).
It is shown that moderate sub-threshold noises
considerably improve the transition fidelity.
We also note that the presence of such noises
does not significantly degrade the transmission 
fidelity for supra-threshold cases in
ensemble neurons.
For a comparison, 
the $g_s$ dependence of $\gamma$ and SNR for single ($N=1$)
neurons is shown in Figs. 11(a) and 11(b), respectively.
As having been shown in Fig. 8(a), SR for $N=1$
is less significant compared to that for $N=500$.

\subsection{Input Pulses with $M=2$ and 3}

Now we discuss the cases of $M=2$ and 3.
Input pulses are applied at $t=100$ and 125 ms for 
the $M=2$ case. The input ISI is assume to be $T_{\rm i}$ = 25 ms
because spikes with this value of ISI are reported to be ubiquitous
in cortical brains \cite{Traub99}.
Firings of 500 neurons for the noise intensity of
$D=1.0$ are shown by raster in Figs. 12(a), which shows that firings
occur mainly at $t \sim 103$ and 128 ms with a delay of
about 3 ms.
The output signal $W_{\rm o}(t)$ averaged
over $N=$ 500 neurons is depicted by the solid curve in
the uppermost frame of Fig. 13(a), which has main two peaks. 
The WT decomposition and WT coefficients of $W_{\rm o}(t)$  are
plotted in Figs. 13(a) and 13(b), respectively.
When we compare the results for $M=2$ shown in Figs. 13(a) and 13(b)
with those for $M=1$ shown in Fig. 5(a) and 5(b), 
we note that components for $j=1$ and 2 for $M=2$ are 
much increased because of
the presence of the second peak for $M=2$ inputs while
the contributions from $j \geq 3$ are changed little.
The denoising has been made by the procedure 
given by Eqs. (37), (39) and (45).
The denoising WT coefficients are plotted by dashed bars in Fig. 13(b)
and the denoising component signals are shown by dashed curves 
in Fig. 13(a). 

We apply the $M=3$ pulse at $t=100$, 125 and 150 ms,
the input ISI being again $T_{\rm i}=25$ ms.
Raster in Fig. 12(b) shows firing of 500 neurons
for $D=1.0$. Firings mainly occur at $t \sim$ 103,
128 and 153 ms with a delay of 3 ms.
The solid curve in the uppermost frame of Fig. 14(a) shows 
the averaged output of $W_{\rm o}(t)$ for $D=1.0$.
The sold bars in Fig. 14(b) show its WT coefficients
and the solid curves in Fig. 14(a) express its WT decomposition. 
Again components of $j=1$ and 2 are increased because of
the second and third peaks in $W_{\rm o}(t)$.
The denoising WT coefficients are shown by dashed bars in Fig. 14(b)
and denoising signals by dashed curves in Fig. 14(a).

The calculated $D$ dependence of the cross-correlation $\gamma$
[SNR] for $M=$ 1, 2 and 3 is plotted 
in Fig. 15(a) [15(b)].
Both $\gamma$ and SNR show typical SR behavior
irrespective of the value of $M$, although
a slight difference exists between the $M$ dependence 
of $\gamma$ and SNR: 
for larger $M$, the former is larger but the latter is smaller
at the moderate noise intensity of $D < 1.0$.  
When similar simulations are performed for different
ISI values of $T_{\rm i}=$ 15 and
35 ms, we obtain results which are almost the
same as that for $T_{\rm i}=$ 25 (not shown).
This is because the output spikes for
inputs with $M=$ 2 and 3 are
superposition of an output spike for a $M=1$ input when the ISI
is larger than the refractory period of neurons.
 


\section{CONCLUSION AND DISCUSSION}


It has been controversial how neurons communicate 
information by action potentials or spikes \cite{Rieke96}-\cite{Pouget00}.
The one issue is whether information is encoded
in the average firing rate of neurons ({\it rate code})
or in the precise firing times ({\it temporal code}).
Since Andrian \cite{Andrian26} first noted the relationship
between neural firing rate and stimulus intensity, 
the rate-code model has been supported in many experiments
of motor and sensory neurons.
In the last several years, however, 
experimental evidences  have been accumulated, 
indicating a  use of
the temporal code in neural systems \cite{Carr86}-\cite{Thorpe96}.
Human visual systems, for example, have shown to classify
patterns within 250 ms despite the fact that at least 
ten synaptic stages are involved from retina to 
the temporal brain \cite{Thorpe96}.
The transmission
times between
two successive stages of synaptic 
transmission are suggested to be no more than 10 ms 
on the average.
This period is too short to allow rates to be determined
accurately.

Although much of debates on the nature of the neural code has focused 
on the rate versus temporal codes, there is the other important
issue to consider: 
information is encoded 
in the activity of single (or very few) neurons
or that of a large number of neurons
({\it population} or {\it ensemble code}).
The population rate code model assumes that information is
coded in the relative firing rates of ensemble neurons,
and has been adopted in the most of the theoretical analysis.
On the contrary, in the population temporal code model,
it is assumed that
relative timings between spikes in ensemble neurons
may be used as an encoding mechanism for perceptional
processing \cite{Hopfield95}-\cite{Rullen01}.
A number of experimental data supporting this code have been reported
in recent years \cite{Gray89}-\cite{Hatso98}.
For example, data have demonstrated that temporally
coordinated spikes can systematically signal sensory
object feature, even in the absence of changes
in firing rate of the spikes \cite{deCharms96}.

Our simulations based on the temporal-code model
has shown that a population of neurons plays a very important
role for the transmission of sub-threshold transient spike signals 
[Figs. 8(a) and 18(b)].
In particular for single neurons the transmission is
quite unreliable and the appreciable SR effect is not realized.
When the size of ensemble neurons is increased,
the transmission fidelity is much improved
in a fairly wide-range of parameters $g_s$
including both the sub- and supra-threshold cases (Fig. 10).
We note in Figs. 8(a) and 8(b) that $\gamma$ (or SNR)
for $N=100$ is different from and larger than 
that for $N=1$ with 100 trials.
This seems strange because a simulation for an ensemble
of $N$ neurons is expected to be equivalent to simulations
for a single neuron with $N$ trials,
if there is no couplings among neurons as in 
our model.
This is, however, not true, and it will be understood as follows.
We consider a quantity of $X(N, N_r)$ which is $\gamma$
(or SNR) averaged over $N_r$ trials for an ensemble of
$N$ neurons. We implicitly express $X(N, N_r)$ as
\begin{eqnarray}
X(N, N_r) &=& \ll F(<w_i^{(\mu)}>) \gg  \\
& = & \frac{1}{N_r}\; \sum_{\mu=1}^{N_r}
F(\frac{1}{N}\; \sum_{i=1}^{N} w_i^{(\mu)}),
\end{eqnarray}
with
\begin{equation}
w_{\rm i}^{(\mu)} = w_{\rm i}^{(\mu)}(t) 
= \sum_n \; \Theta(\mid t - t_{{\rm o}in}^{(\mu)}\mid - T_{\rm b}/2),
\end{equation}
where $\ll \cdot \gg$ and $<\cdot>$ stand for averages over trials and
an ensemble neurons, respectively, defined by Eq.(47),
$t_{{\rm o}in}^{(\mu)}$ is the $n$-th firing time of the neuron $i$
in the $\mu$-th trial, $w_{\rm i}^{(\mu)}(t)$ is its output 
signal of the neuron $i$
within a time bin of $T_{\rm b}$ [{\it cf.} Eq.(29)], 
and $F(y(t))$ is a functional of a given function of $y(t)$
relevant to a calculation of $\gamma$ (or SNR).
Figures 8(a) and 8(b) show that the relation:
$X(100, 1) > X(1, 100)$, namely
\begin{equation}
F(<w_i^{(1)}>) \; > \; \ll F(w_{1}^{(\mu)}) \gg,
\end{equation}
holds for our $\gamma$ and SNR.
Note that if $F(\cdot)$ is linear, we get $X(100, 1) = X(1,100)$. 
This implies that the inequality given by Eq. (49) 
is expected to arise from a {\it non-linear} character
of $F(\cdot)$. 
This reminds us of the algebraic inequality:
$f(<x>) \;\geq \;<f(x)>$ valid
for a convex function $f(x)$, where
the bracket $<\cdot>$ stands for an average over a distribution of 
a variable $x$.
It should be again noted that there is no couplings among our neurons in 
the adopted model.
Then the enhancement of SNR with increasing $N$ is only
due to a population of neurons.
This is quite different from the result of some papers
\cite{Kanamaru01}\cite{Stocks01}\cite{Wang00}\cite{Liu01}
in which the transmission fidelity is enhanced
not only by noises but also by introduced couplings
among neurons in an ensemble.

In our simulations reported in Sec. III, 
independent noises are applied to ensemble neurons. 
If instead we try to apply the same or {\it completely correlated} noise
to them, it is equivalent to applying noises to a single
neuron, and then appreciable SR effect is not realized
as discussed above.  Then SR for transient spikes
requires independent noises
to be applied to a large-scale ensemble of neurons.
This is consistent with the result of Liu, Hu and Wang \cite{Liu01}
who discussed the effect of correlated noises on SR for stationary
analog inputs.

Although spike trains with small values of $M=1-3$ 
have been examined in Sec. III,
we can make an analysis of spikes with larger $M$
or bursts, by using our method.
In order to demonstrate its feasibility,
we have made simulations for transient spikes with larger $M$.
The upper curve of Fig. 16(a) expresses  
input $W_{\rm i}(t)$ with a $M=7$ spike train whose firing times are
$t_{{\rm i}m}$= 100, 115, 130, 145, 160, 180, and 200 ms,
and ISIs are $T_{\rm i}$ =15 and 20 ms. 
Firings of 100 neurons in an ensemble is depicted by raster in Fig. 16(b),
where the parameters of $g_s=0.06$, $D=1.0$ and $N=100$ are adopted.
The lower curve in Fig. 16(a) shows its output $W_{\rm o}(t)$
averaged over the ensemble.
We apply the WT to $W_{\rm o}(t)$ to get its WT coefficients
and its WT decomposition, the latter being shown in Fig. 16(b).
The $j=1$ and $j=2$ components are dominant.
After the denoising, we get $\gamma=0.523$ and $SNR$=18.6 dB,
which are comparable to those for $D=1.0$ with $M=1-3$ 
shown in Figs. 15(a) and 15(b).
In our denosiging method given by Eqs.(39) and (45), we extract noises
outside the $b$ region relevant to a cluster of spikes, but
do not take account of noises between pulses.  
When a number of pulses $M$ and/or
the ISI $T_{\rm i}$ become larger, 
a contribution from noises between pulses
become considerable, and it is necessary to modify the denoising
method such as to extract noises between pulses, 
for example, as given by
\begin{eqnarray}
c_{jk}^{dn} & = & c_{jk}, \;\;\;
\mbox{for $j \geq j_c$ or $k_{{\rm L}m} \leq k \leq k_{{\rm U}m}$} 
\;\; \mbox{($m=1-M$)} \nonumber \\
& = & 0.  \;\;\;\;\; \mbox{otherwise}
\end{eqnarray}
In Eq. (50) $k_{Lm}$ and $k_{Um}$ are $m$- and $j$-dependent 
lower and upper limits
given by
\begin{equation}
k_{\rm L m}= 2^{-j} \;(t_{{\rm o}m}/T_{\rm b}-\delta b), \;\;\;\;
k_{\rm U m}=2^{-j} \;(t_{{\rm o}m}/T_{\rm b}+\delta b),
\end{equation}
where $t_{{\rm o}m}$ is the $m$-th firing time for the noise-free and 
supra-threshold input and $\delta b$ the margin of $b$.

An ensemble of neurons adopted in our model can be regarded 
as the front layer (referred to as the layer I here)
of a synfire chain \cite{Abeles93}: output spikes of the layer I are
feed-forwarded to neurons on the next layer (referred to as the layer II)
and spikes propagate through the synfire chain. 
The postsynaptic current of a neuron $\ell$ on the layer II is
given by
\begin{equation}
I_{\ell}^{\rm ps} = 
(1/N) \:\sum_{i=1}^N \sum_{m} \: w_{\ell i}\: (V_a - V_s) 
\: \alpha(t-\tau_{\ell i}-t_{{\rm o}i m})
+ I_{\ell}^{\rm n}, 
\end{equation}
where $w_{\ell i}$ and $\tau_{\ell i}$ are the synaptic 
coupling and delay time, respectively, for spikes to
propagate from neuron $i$ on the layer I 
to neuron $\ell$ on layer II,
$t_{{\rm o}i m}$ the $m$-th firing time of neuron $i$
on the layer I, 
and $I_{\ell}^{\rm n}$ noises.
Transmission of spikes through the synfire chain 
depends on the couplings (excitatory or inhibitory) $w_{\ell i}$,
the delay time $\tau_{\ell i}$ and noises $I_{\ell}^{\rm n}$.
There are some discussions on the stability of
spike transmission in a synfire chain.
Quite recently Diesmann \cite{Diesmann99} has shown
by simulations using IF models
that the spike propagation with a precise timing
is possible in fairly large-scale synfire chains
with moderate noises.

By augmenting our neuron model including
the coupling term given by Eq.(52) and
tentatively assuming $w_{\ell i}=w=1.5$ and $I_{\ell}^{\rm n}=0$, 
we have calculated the transmission fidelity by applying
our WT analysis to output signals on the layer II
as well as those on the layer I.
Calculated cross-correlation and SNR are shown 
in Fig. 17(a) and 17(b), respectively.
In Fig. 17(a), $\gamma_{L}$ (L=I, II) denotes the
cross-correlation between the input and output signals
on the layer $L$. Similarly, $SNR_{L}$ in Fig. 17(b)
expresses the SNR of output signal on the layer $L$.
Note that $\gamma_{I}$ and $SNR_{I}$ are nothing but
$\gamma$ and SNR having been shown in Figs. 6(a) and 6(b).
We note that the transmission fidelity on the layer II
is better that that of the layer I
because $\gamma_{I} < \gamma_{II}$ and $SNR_{I} < SNR_{II}$
at almost $D$ values.
We have chosen $w=1.5$ such that neurons on the layer II
can fire when more than 6\%\ of neurons on the layer I fire.
When we adopt smaller (larger) value of $w$,
both $\gamma_{II}$ and $SNR_{II}$ abruptly increase
at larger (smaller) $D$ value.  However,
the general behavior of the $D$
dependence of $\gamma_{II}$ and $SNR_{II}$ is not changed.
This improvement of the transmission fidelity in the layer II
than in the front layer I
is expected to be partly due to the threshold-crossing 
behavior of HH neurons. 
It would be interesting to investigate the
transmission of signals in a synfire chain
by including SR of its front layer.

Our paper entirely relies on numerical simulations.
We are currently trying to work on theoretical description of the
result reported in this paper.
Conventional approaches having been employed for 
a study of SR such as the rate-equation and 
linear-response theories \cite{Gammai98}-\cite{Anish99},
do not work on our case.
Mato \cite{Mato98} adopted Gammaitoni's approach \cite{Gammaitoni95}
for an analysis of his SR result
with the continuous spike-train signals.
It seems, however, not to be transposed directly
to our case of transient spike-train signals 
even if our HH model is replaced by simpler IF model
or threshold-crossing model.
Its analytical study is left as our future problem.

To summarize, 
the response of an ensemble of neurons 
to transient spike trains has been discussed
with the use of the WT.
Although the transmission of sub-threshold transient inputs
is not reliable in a single neuron, it is much improved
in a large-scale ensemble of neurons with a weak noise (SR effect).
It is also shown that
this noise does not significantly degrade the transmission
fidelity for supra-threshold inputs in the large-scale ensemble.
A population of neurons plays an 
essential role in SR for transient inputs.

\section*{Acknowledgements}
This work is partly supported by
a Grant-in-Aid for Scientific Research from the Japanese 
Ministry of Education, Culture, Sports, Science and Technology.



\begin{figure}
\caption{
The time dependence of 
(a) the $M=1$ input pulse ($U_{\rm i}$), 
(b) the postsynaptic current 
($I_{1}=I_{1}^{\rm ps}+I_{1}^{\rm n}$),
(c) the membrane potential ($V_{1}$), and
(d) input ($W_{\rm i}$) and 
(e) output signals ($W_{\rm o}$) within
the time bin of $T_{\rm b}=5$ ms for $D=0.5$, 
$g_s=0.06$ and $N=500$.
}
\label{fig1}
\end{figure}

\begin{figure}
\caption{
Rasters showing firings in ensemble neurons
for (a) $D=0.5$ and (b) $D=1.0$ with 
$M=1$, $g_s=0.06$ and $N=500$.
}
\label{fig2}
\end{figure}

\begin{figure}
\caption{
(a) The input signal $W_{\rm i}(t)$ (uppermost frame) 
for $M=1$, and
its WT decomposition: $f=\sum_{j=1}^{5} \; f_j$.
(b) Its WT expansion coefficients $c_{jk}$.
Curves of $f_{j}$ for $j \geq 2$ and of $f$ in (a), and 
those of $c_{jk}$ for $j \geq 2$ in (b) are successively
shifted upward by 1.0.
Upper horizontal scale expresses $b\; T_{\rm b}$ in units of ms.
}
\label{fig3}
\end{figure}

\begin{figure}
\caption{
(a) The output signal $W_{\rm o}(t)$ (uppermost frame) 
for $D=0.5$ and $M=1$, and
its WT decomposition: $f=\sum_{j=1}^{5} \; f_j$.
(b) Its WT expansion coefficients $c_{jk}$.
Solid and dashed curves denote
the original and denoising results, respectively.
Same as in Fig. 3.
}
\label{fig4}
\end{figure}

\begin{figure}
\caption{
(a) The output signal $W_{\rm o}(t)$ (uppermost frame) 
for $D=1.0$ and $M=1$, and
its WT decomposition: $f=\sum_{j=1}^{5} \; f_j$.
(b) Its WT expansion coefficients $c_{jk}$.
Solid and dashed curves denote
the original and denoising results, respectively.
Same as in Fig. 3.
}
\label{fig5}
\end{figure}

\begin{figure}
\caption{
The $D$ dependence of
$\gamma$ (circles), $\Gamma_{\rm oo}$ (triangles)
and $\Gamma$ (squares)
for $M=1$, $g_s=0.06$ and $N=500$. 
}
\label{fig6}
\end{figure}

\begin{figure}
\caption{
The $\delta b$ dependence 
of SNR of output signals for (a) $D=0.5$ and (b) $D=1.0$ 
with $j_c=2$ (circles), 3 (triangles) and 4 (squares)
for $M=1$, $g_s=0.06$ and $N=500$,
the denoising condition being given by Eqs.(39) and (45).
}
\label{fig7}
\end{figure}

\begin{figure}
\caption{
(a) The cross-correlation $\gamma$ and (b) SNR as 
a function of $D$ for $M=1$ and $g_s=0.06$ with $N=$ 1, 10, 100 and 500.
Results for $N=1$ ($N=10$) are averaged values of
100 (10) trials and their RMS are shown by error bars.
Results of (a) and (b) for $N=$ 10, 100 and 500
are successively shifted upward by 1.0 and 20, respectively.
}
\label{fig8}
\end{figure}

\begin{figure}
\caption{
(a) The cross-correlation $\gamma$ and (b) SNR
of 100 trials for single ($N=1$) neurons with $g_s=0.06$ and $M=1$. 
}
\label{fig9}
\end{figure}

\begin{figure}
\caption{
The $g_s$ dependence of the cross-correlation 
$\gamma$ and SNR for $N=500$ and $M=1$,
Dot, dashed and solid curves denoting the results
of $D=0.0$, 0.5 and 1.0, respectively.
The arrow expresses the threshold $g_s$ value.
}
\label{fig10}
\end{figure}

\begin{figure}
\caption{
The $g_s$ dependence of the cross-correlation 
$\gamma$ and SNR for $N=1$ and $M=1$:
Dot, dashed and solid curves denote the results
of $D=0.0$, 0.5 and 1.0, respectively, and
error bars express RMS values for 100 trials.
}
\label{fig11}
\end{figure}

\begin{figure}
\caption{
Raster showing firings in ensemble neurons
for (a) $M=2$ and (b) $M=3$ with 
$D=1.0$, $g_s=0.06$ and $N=500$.
}
\label{fig12}
\end{figure}

\begin{figure}
\caption{
(a) The output signal $W_{\rm o}(t)$ (uppermost frame) 
for $D=1.0$ and $M=2$, and
its WT decomposition: $f=\sum_{j=1}^{5} \; f_j$.
(b) Its WT expansion coefficients $c_{jk}$.
Solid and dashed curves denote
the original and denoising results, respectively.
Same as in Fig. 3.
}
\label{fig13}
\end{figure}

\begin{figure}
\caption{
(a) The output signal $W_{\rm o}(t)$ (uppermost frame) 
for $D=1.0$ and $M=3$, and
its WT decomposition: $f=\sum_{j=1}^{5} \; f_j$.
(b) Its WT expansion coefficients $c_{jk}$.
Solid and dashed curves denote
the original and denoising results, respectively.
Same as in Fig. 3.
}
\label{fig14}
\end{figure}

\begin{figure}
\caption{
(a) The cross-correlation $\gamma$ and (b) SNR as 
a function of $D$ for $M=1$ (circles), 
$M=2$ (squares) and $M=3$ (triangles).
}
\label{fig15}
\end{figure}

\begin{figure}
\caption{
(a) The input spike train $W_{\rm i}(t)$ (upper curve)
consisting of $M=10$ pulses, 
and its output signal $W_{\rm o}(t)$ (lower curve) for
$g_s=0.06$, $D=1.0$ and $N=100$.
(b) Raster showing firings of 100 neurons.
(c) The WT decomposition of $W_{\rm o}(t)$,
results for $j=$2, 3, 4 and 5 being successively 
shifted upward by 0.2.
}
\label{fig16}
\end{figure}

\begin{figure}
\caption{
(a) The $D$ dependence of the cross-correlation
$\gamma_{I}$ (filled circles) 
and $\gamma_{II}$ (open circles), and 
(b) that of $SNR_{I}$ (filled circles)
and $SNR_{II}$ (open circles) for $M=1$; note that
$SNR_{II} = \infty$ for $0.3 < D < 0.6$ (see text).
}
\label{fig17}
\end{figure}


\begin{thebibliography}{99}
\bibitem{Gammai98}L. Gammaitoni, P. H\"{a}nnggi, P. Jung, 
and F. Marchesoni,
Rev. Mod. Phys. {\bf 70}, 223 (1998).

\bibitem{Anish99}V. S. Anishchenko, A. B. Neiman, F. Moss
and L. Schimansky-Geier,
Soviet Phys.-Uspekhi {\bf 42}, 7 (1999).


\bibitem{Bulsara96}A. R. Bulsara, T. C. Elston, C. R. Doering,
and K. Lindenberg,
Phys. Rev. E {\bf 53}, 3958 (1996).

\bibitem{Plesser97}H. E. Plesser and S. Tanaka,
Phys. Lett. A  {\bf 225}, 228 (1994).


\bibitem{Shimokawa99a}T. Shimokawa, K. Pakdaman, and S. Sato,
Phys. Rev. E {\bf 59}, 3427 (1999).

\bibitem{Longtin93}A. Longtin,
J. Stat. Phys. {\bf 70}, 309 (1993).

\bibitem{Wiesenfeld94}K. Wiesenfeld, D. Pierson, E. Pantazelou,
C. Dames, and F. Moss,
Phys. Rev. Lett. {\bf 72}, 2125 (1994).

\bibitem{Longtin94}A. Longtin and D. R. Chialvo,
Phys. Rev. Lett. {\bf 81}, 4012 (1994).

\bibitem{Lee99}S. Lee and S. Kim,
Phys. Rev. E {\bf 60}, 826 (1999).

\bibitem{Shimokawa99b}T. Shimokawa, A. Rogel, K. Pakdaman, and S. Sato,
Phys. Rev. E {\bf 59}, 3461 (1999).

\bibitem{Lindner01}B. Lindner and L. Schimansky-Geier,
Phys. Rev. Lett. {\bf 86}, 2934 (2001).


\bibitem{Collins95a}J. J. Collins, C. C. Chow and T. T. Imhoff,
Nature  {\bf 376}, 236 (1995).

\bibitem{Kanamaru01}T. Kanamaru, T. Horita, and Y. Okabe,
Phys. Rev. E {\bf 64}, 31908 (2000).

\bibitem{Stocks01}N. G. Stocks, and R. Mannella,
Phys. Rev. E {\bf 64}, 30902 (2001).


\bibitem{Pei96a}X. Pei, L. Wilkens, and F. Moss,
Phys. Rev. Lett. {\bf 77}, 4679 (1996).

\bibitem{Tanabe99}S. Tanabe, S. Sato, and K. Pakdaman,
Phys. Rev. E {\bf 60}, 7235 (1999).

\bibitem{Wang00}Y. Wang, D. T. W. Chik, and Z. D. Wang,
Phys. Rev. E {\bf 61}, 740 (2000).

\bibitem{Liu01}F. Liu, B. Hu, and W. Wang,
Phys. Rev. E {\bf 63}, 31907 (2000).

\bibitem{Douglass93}J. K. Douglass, L. Wilkens, E. Pantazelou,
and F. Moss,
Nature  {\bf 365}, 337 (1993).

\bibitem{Pei96b}X. Pei, L. A. Wilkens, and F. Moss,
J. Neurophysiol. {\bf 76}, 3002 (1996).

\bibitem{Levins96}J. E. Levins and J. P. Miller,
Nature {\bf 380}, 165 (1996).

\bibitem{Gluckman96}B. J. Gluckman, J. I. Netoff, E. J. Neel,
W. L. Ditto, M. L. Spano, and S. J. Shiff,
Phys. Rev. Lett. {\bf 77}, 4098 (1996).

\bibitem{Nozaki99}D. Nozaki, D. J. Mar, P. Grigg, and J. J. Collins,
Phys. Rev. Lett. {\bf 82}, 2402 (1999).

\bibitem{Chapeau96}F. Chapeau-Blondeau, X. Godivier, and
N. Chambet,
Phys. Rev. E {\bf 53}, 1273 (1996).

\bibitem{Deco98}G. Deco and B. Sch\"{u}rmann,
Physica D {\bf 117}, 276 (1998).

\bibitem{Mato98}G. Mato,
Phys. Rev. E {\bf 58}, 876 (1998);
{\it ibid.} {\bf 59}, 3339 (1999).


\bibitem{Collins95b}J. J. Collins, C. C. Chow and T. T. Imhoff,
Phys. Rev. E {\bf 52}, R3321 (1995);
{ibid.} {\bf 54}, 5575 (1996).

\bibitem{Fakir98}R. Fakir,
Phys. Rev. E {\bf 57},  6996(1998);
{\it ibid.} {\bf 58}, 5175 (1998).





\bibitem{Traub99}R. D. Traub, J. G. R. Jefferys and
M. A. Whittington:
{\it Fast Oscillations in Cortical Circuits} 
(MIT press, Cambridge, 1999).

\bibitem{Sullivan98}W. E. Sullivan and M. Konishiki:
J. Nerosci. {\bf 393}, 268 (1998).

\bibitem{Rose85}G. Rose and W. Heilingenberg:
Nature {\bf 318}, 178 (1985).

\bibitem{Astaf96}For reviews on the WT and
its application see; N. M. Astaf'eva:
Physics-Usp. {\bf 39}, 1085 (1996);
I. M. Dremin, O. V. Ivanov, and V. A. Nechitailo:
hep-ph/0101182.

\bibitem{Samar99}V. J. Samar, A. Bopardikar, R. Rao and K. Swartz:
Brain and Language {\bf 66}, 7 (1999).


\bibitem{Blanco96}S. Blanco, C. E. D'Attellis, S. I. Isaacson,
O. A. Rosso, and R. O. Sirne:
Phys. Rev. E {\bf 54}, 6661 (1996).

\bibitem{Blanco98}S. Blanco, A. Figliola, R. Q. Quiroga, O. A. Rosso,
and E. Serrano:
Phys. Rev. E {\bf 57}, 932 (1998).

\bibitem{Ratz99}J. Ratz, L. Dickerson, and B. Turetsky:
Brain and Language {\bf 66}, 61 (1999).

\bibitem{Trejo99}L. J. Trejo, and M. J. Shensa:
Brain and Language {\bf 66}, 89 (1999).

\bibitem{Demiralp99a}T. Demiralp, A. Ademoglu, M. Sch\"urmann,
C. Ba\c sar-Eroglu, and E. Ba\c sar:
Brain and Language {\bf 66}, 108 (1999).

\bibitem{Demiralp99b}T. Demiralp, J. Yordanova, V. Kolev,
A. Ademoglu, M. Devrin, and V. J. Samar:
Brain and Language {\bf 66}, 29 (1999).


\bibitem{Rosso01}O. A. Rosso, S. Blanco, J. Yordanova, 
V. Kolev, A. Figliola, M. Sch\"urmann, E. Ba\c sar:
J. Neurosci. Methods {\bf 105}, 65 (2001).

\bibitem{Hulata00}E. Hulata, R. Segev, Y. Shapir, M. Benveniste,
and E.Ben-Jacob:
Phys. Rev. Lett. {\bf 85}, 4637 (2000).

\bibitem{Letelier00}J. C. Letelier and P. P. Weber:
J. Neurosci. Methods {\bf 101}, 93 (2000).

\bibitem{Zour97}G. Zouridakis, and D. C. Tam:
Comput. Biol. Methods {\bf 27}, 9 (1997).


\bibitem{Strat01}D. Stratimirovi\'c, S. Milosevi\'c, S. Blesi\'c
and M.Ljubiavljevi\'c:
Physica A {\bf 291}, 13 (2001).


\bibitem{Hasegawa01}H. Hasegawa,
E-print: cond-mat/0109444.


\bibitem{Destexhe98}A. Destexhe, Z. F. Mainen, and T. J. Sejnowski,
in {\it The Handbook of Brain Theory and Neural Networks'},
ed. M. A. Arbib. MIT press, Cambridgr 1998, p956

\bibitem{Smith98}R. G. Smith,
in {\it The Handbook of Brain Theory and Neural Networks'},
ed. M. A. Arbib. MIT press, Cambridgr 1998, p816

\bibitem{Shadlen94}M. N. Shadlen and W. T. Newsome,
Curr. Opin.Nurobiol. {\bf 4}, 569 (1994).

\bibitem{Hasegawa02}H. Hasegawa, (unpublished).

\bibitem{Hodgkin52}A. L. Hodgkin and A. F. Huxley,
J. Physiol. {\bf 117}, 500 (1952).


\bibitem{Bartnik92}E. A. Bartnik, K. J. Blinowska, and P. J. Durks,
Biol. Cybem. {\bf 67}, 175 (1992).

\bibitem{Bertrand94}O. Bertrand, J. Bohorquez, and J. Pemier,
IEEE Trans, Biomed, Eng. {\bf 41}, 77 (1994).

\bibitem{Donoho95}D. L. Donoho, I. M. Johnstone, and B. W. Silverman,
IEEE Trans, Inform. Theory {\bf 41}, 613 (1995).

\bibitem{Quiroga00}R. Q. Quiroga,
Physica D {\bf 145}, 278 (2000).

\bibitem{Stocks00}N. G. Stocks,
Phys. Rev. Lett. {\bf 84}, 2310 (2000);
Phys. Rev. E {\bf 63}, 41114 (2001).

\bibitem{Hasegawa00}H. Hasegawa,
Phys. Rev. E {\bf 61}, 718 (2000);
{\it ibid.}  {\bf 62}, 1456 [E];
Bull. Tokyo Gakugei Univ. Ser. 4, {\bf 53}, 31 (2001).

\bibitem{Mainen95}Z. F. Mainen, and T. J. Sejnowsky,
Reliability of spike timing in neocortical neirons
Science  {\bf 268}, 1503 (1995).


\bibitem{Rieke96}F. Rieke, D. Warland, R. Steveninck and W. Bialek:
{\it Exploring the Neural Code} (MIT press, Cambridge, 1996).

\bibitem{deCharms98}R. C. deCharms: 
Proc. Natl. Acad. Sci USA {\bf 95}, 15166 (1998).

\bibitem{Eggermont98}J. J. Eggermont:
Neurosci. Biobehav. Rev, {\bf 22}, 355 (1998).

\bibitem{Ursey99}W. M. Ursey and R. C. Reid:
Annu. Rev. Physiol. {\bf 61}, 435 (1999).

\bibitem{deCharms00}R. C. deCharms and A. Zador:
Ann. Rev. Neurosci. {\bf 23}, 613 (2000).

\bibitem{Pouget00}A. Pouget, P. Dayan and R. Zemel:
Nature Neurosci. {\bf 1}, 125 (2000).

\bibitem{Andrian26}A. D. Andrian,
J. Physiol. (London)  {\bf 61}, 49 (1926).

\bibitem{Carr86}C. E. Carr, W. Heiligenberg
and G. J. Rose,
J. Neurosci. {\bf 6}, 107 (1986).

\bibitem{Eckhorn88}R. Eckhorn, R. Bauer, W. Jordan,
M. Brosch, W. Kruse, M. Munk, and H. J. Reitboeck,
Biol. Cybern. {\bf 60} 121, (1988)

\bibitem{Gray89}C. M. Gray and W. Singer,
Proc. Natl. Acad. Sci. (USA)  {\bf 86}, 1698 (1989).

\bibitem{Rolls94}E. T. Rolls and M. J. Tovee,
Proc. Roy. Soc. B {\bf 257} 9, (1994)

\bibitem{Thorpe96}S. Thorpe, D. Fize and C. Marlot,
Nature (London) {\bf 381}, 520 (1996).


\bibitem{Hopfield95}J J Hopfield:  Nature {\bf 376}, 33 (1995).

\bibitem{Horn98}D. Horn and S. Levanda:
Neural Comput. {\bf 10}, 1705 (1998). 

\bibitem{Rullen01}R. van Rullen and S. J. Thorpe:
Neural Comput. {\bf 13}, 1255 (2001).

\bibitem{Gray89}C. M. Gray and W. Singer:
Proc. Natl. Acad. Sci. USA {\bf 86}, 1698 (1989).

\bibitem{deCharms96}R. C. deCharmes and M. M. Merzenich:
Nature {\bf 381}, 610 (1996).

\bibitem{Traub97}R. D. Traub, M. A. Whittington, 
J. G. R. Jefferys:
Neural Comput. {\bf 9}, 1251 (1997).

\bibitem{Hatso98}N. Hatsopoulas, C. L. Ojakangas, L. Paninski
and J. P. Donohue:
Proc. Natl. Acad. Sci. USA {\bf 95}, 15706 (1998).


\bibitem{Abeles93}M. Abeles, H. Bergman, E. Margalit, 
and E. Vaadia, 
J. Neurophys. {\bf 70}, 1629 (1993).

\bibitem{Diesmann99}M. Diesmann, M. Gewaltig, and
A. Aertsen,
Nature  {\bf 402}, 529 (1999).


\bibitem{Gammaitoni95}L. Gammaitoni,
Phys. Rev. E {\bf 52}, 4691 (1995).



\end{thebibliography}
\end{document}